\documentclass{aastex}
\usepackage{spr-astr-addons}
\usepackage{url}
\usepackage{subfigure}

\RequirePackage{color}

\begin{document}

\title{Building the cosmic distance scale: \\ from Hipparcos to Gaia}
\shorttitle{Building the cosmic distance scale: from Hipparcos to Gaia}
\shortauthors{Turon, Luri  \& Masana}

\author{Catherine Turon}
\altaffiltext{}{GEPI - UMR 8111, Observatoire de Paris, CNRS, Universit\'e Paris Diderot, 5 Place Jules Janssen, 92190 Meudon, France, catherine.turon at obspm.fr}

\and \author{Xavier Luri} 
\and \author{Eduard Masana}
\altaffiltext{}{Departament d'Astronomia i Meteorologia (ICCUB-IEEC), Universitat de Barcelona, C/ Mart\'{\i} i Franqu\`es 1, 08028, Barcelona, Spain, xluri at am.ub.es \altaffilmark{2}}

\begin{abstract}

Hipparcos, the first ever experiment of global astrometry, was launched by ESA (European Space Agency) in 1989 and its results published in 1997 (Perryman et al., Astron. Astrophys. 323, L49, 1997; Perryman \& ESA (eds), The Hipparcos and Tycho catalogues, ESA SP-1200, 1997). A new reduction was later performed using an improved satellite attitude reconstruction leading to an improved accuracy for stars brighter than 9$^{th}$ magnitude (van Leeuwen \& Fantino, Astron. Astrophys. 439, 791, 2005; van Leeuwen, Astron. Astrophys. 474, 653, 2007).
The Hipparcos Catalogue provided an extended dataset of very accurate astrometric data (positions, trigonometric parallaxes and proper motions), enlarging by two orders of magnitude the quantity and quality of distance determinations and luminosity calibrations. The availability of more than 20\,000 stars (22\,000 for the original catalogue, 30\,000 for the re-reduction) with a trigonometric parallax known to better than 10\,\% opened the way to a drastic revision of our \mbox{3-D} knowledge of the solar neighbourhood and to a renewal of the calibration of many distance indicators and age estimations. The prospects opened by Gaia, the next ESA cornerstone, planned for launch in June 2013 (Perryman et al., Astron. Astrophys. 369, 339, 2001), are still much more dramatic: a billion objects with systematic and quasi simultaneous astrometric, spectrophotometric and spectroscopic observations, about 150 million stars with expected distances to better than 10\,\%, all over the Galaxy. All stellar distance indicators, in very large numbers, will be directly measured, providing a direct calibration of their luminosity and making possible detailed studies of the impacts of various effects linked to chemical element abundances, age or cluster membership. With the help of simulations of the data expected from Gaia, obtained from the mission simulator developed by DPAC (Gaia Data Processing and Analysis Consortium), we will illustrate what Gaia can provide with some selected examples.
\end{abstract}

\keywords{Space observatory; Astrometry; Hipparcos; Gaia; stars: distances; stars: fundamental parameters; star clusters; pulsating variable stars; distance scale.}

\section{Introduction}
The last 15 years have seen some major improvements in the determination of the fundamental distance scale, due mainly to three factors: the first ever availability of a large number of precise trigonometric parallaxes from Hipparcos, 
a big effort in obtaining spectroscopic and photometric observations of various stellar candles to study the effects of  colour, metallicity, age, cluster membership, etc. on the absolute luminosity and, finally, the observation of further and further stellar candles in external resolved galaxies with the Hubble Space Telescope and with large telescopes on the ground. 

However, even though Hipparcos was a major improvement with respect to earlier ground-based astrometric observations, {\it only} about 30\,000 stars (compared to a few hundreds before it) were observed with a relative accuracy on their trigonometric parallax better than 10\,\%. Furthermore, all of them are in the solar neighbourhood and very few standard candles are among them. On the other hand, many different photometric and spectroscopic systems have been used to work with these stars, resulting in the non-uniformity of the colour and/or abundances scales, and they have been compared with many different models of stellar atmospheres, resulting in a variety of transformations from colour to effective temperature and of estimations of the bolometric correction. As a result, it is difficult to safely compare observations between themselves and with theoretical isochrones.

In this context, the remaining major sources of uncertainty are: 

\begin{itemize}

  \item The location of the principal sequences of the Hertz-sprung-Russell diagram 
        (main sequence, subgiant branch, turn-off stars, red clump stars, blue supergiants) 
        versus metallicity, age or detailed element abundances,

  \item The calibration of the period-luminosity(-colour) relations of pulsating variable 
        stars with respect to all effects likely to affect their absolute luminosity,
        
  \item The distance (and depth) of the Large Magellanic Cloud, whose Cepheids are often 
        used as reference to derive relative distances to other galaxies.

\end{itemize}

The coming decade should see major improvements in all these aspects: Gaia \citep{Lindegren_Perryman96, Perryman01, Lindegren10} will bring a huge amount of extremely accurate trigonometric parallaxes for very large samples of all galactic populations, the direct distance determination of large samples of all kinds of stellar candles and may even provide the first direct test of the universality of the period-luminosity(-colour) relations. It will also provide a systematic diagnostic of the duplicity (multiplicity) of all observed targets and reliable abundances (and ages from a comparison to isochrones) for very large samples of field and cluster stars. For a more detailed presentation of Gaia, see, among other references,  \cite{Perryman01},  \cite{Lindegren08} or \cite{ELSA2010}. See also papers by \cite{deBruijne11a} and \cite{Eyer11} in this volume.

Besides Gaia, the Hubble Space Telescope (HST), after the major effort put on the HST Key Project on the extragalactic distance scale  \citep{Freedman01}, continues to provide trigonometric parallaxes for targeted stellar candles \citep{Benedict07, McArthur11}. On the ground, a number of large photometric and spectroscopic surveys are under way or planned in the coming years, some of them directly aiming at the improved calibration of the local extragalactic distance scale \citep[Araucaria project,][]{Gieren01}, others to complement Gaia with high-resolution spectroscopy (Hermes project: \citeauthor{Freeman10} \citeyear{Freeman10}, Gaia-ESO survey: \citeauthor{Gilmore11} \citeyear{Gilmore11}), others more generally to improve the knowledge on element abundances in various types of stars and the understanding of Galactic chemical evolution (for example, RAVE: \citeauthor{Siebert11} \citeyear{Siebert11}, or APOGEE: \citeauthor{Allende08} \citeyear{Allende08}, part of the Sloan Digital Sky Survey). Future huge photometric surveys such as Pan-STARSS \citep{Pan-STARRS} or LSST \citep[Large Synoptic Survey Telescope:][]{lsst, Juric11} will also have major synergies with more dedicated programmes. Finally, much effort is being devoted into theoretical work and modelling of stellar atmospheres.

\section{Gaia simulations} \label{simulations}

Gaia will acquire an enormous quantity of complex and extremely precise
data that will be transmitted daily to a ground station. By the end
of Gaia\textquoteright{}s operational life, around 150 terabytes ($10^{14}$
bytes) will have been transmitted to Earth: some 1000 times the raw
volume from the related Hipparcos mission.

An extensive and sophisticated data processing mechanism is being
developed to yield meaningful results from collected data \citep{Mignard08,Mignard11}. To allow
its development and testing, a system has been developed to generate 
the simulated Gaia data, {\bf the Gaia simulator}.

The Gaia simulator has been organised around a common tool box (named
GaiaSimu library) containing two main modules: the instrument model and
the Universe model. The first one describes all the physics of the 
Gaia instruments, from the first reflexion of the light by the mirrors 
to the compression of the telemetry data to be sent to ground. The second
one allows the simulation of the characteristics of all the different types 
of objects that Gaia will observe: their spatial distribution, photometry, 
kinematics and spectra. The Universe model is designed to generate lists 
of astronomical objects whose distribution and the statistics of their observables
are as realistic as possible. It includes not only the simulation of Milky 
Way stars, but also solar system objects, exoplanets, external galaxies and QSOs.

We refer the reader to \cite{Robin11} for a detailed description of the 
model and suffice it to say here that it is based on the Besan\c{c}on Galaxy Model
(BGM) of stellar population synthesis of the Galaxy \citep{Robin03} which provides the distribution of the stars, 
their intrinsic parameters and their motions, taking into account the 3-D Galactic extinction model developed by \cite{Drimmel03}. We present here a short 
summary of its results.

The Universe model generates a total
number of one billion galactic objects with $G<20$ of which $\sim$49\% are
single stars and $\sim$51\% stellar systems formed by stars with
planets and binary/multiple stars.
Individually, the model has created 1\,600\,000\,000 stars
where 31.66\% of them are single stars with magnitude G
inferior to 20 (potentially observable by Gaia) and 68.34\%
correspond to stars in multiple systems. This last
group is formed by stars that have magnitude G inferior to 20
as a system but, in some cases, its isolated components can have
magnitude G superior to 20 and will not be individually detectable
by Gaia.

\begin{figure*}[ht!]
 \centering
 \includegraphics[width=0.495\textwidth,clip]{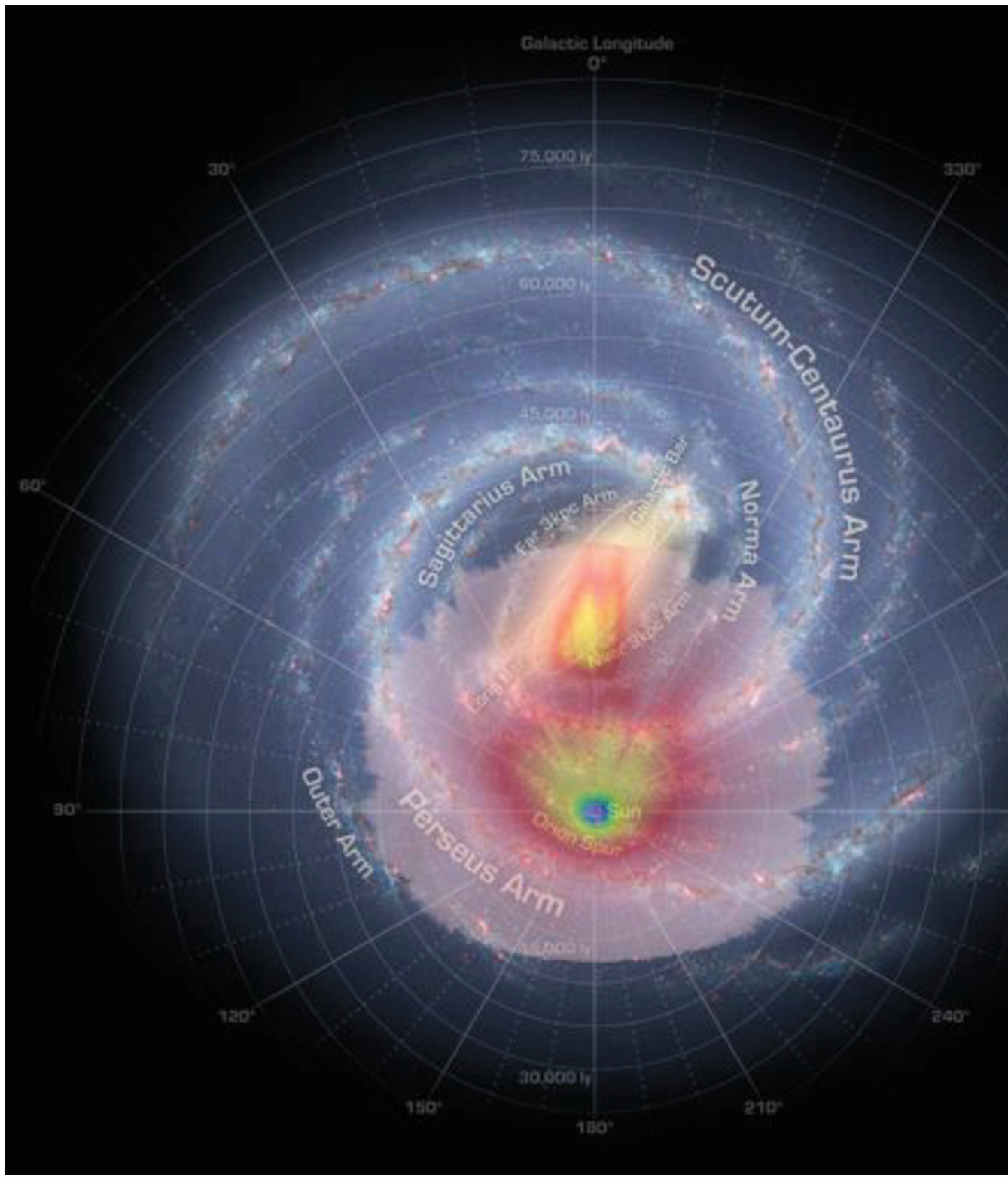}  
 \includegraphics[width=0.495\textwidth,clip]{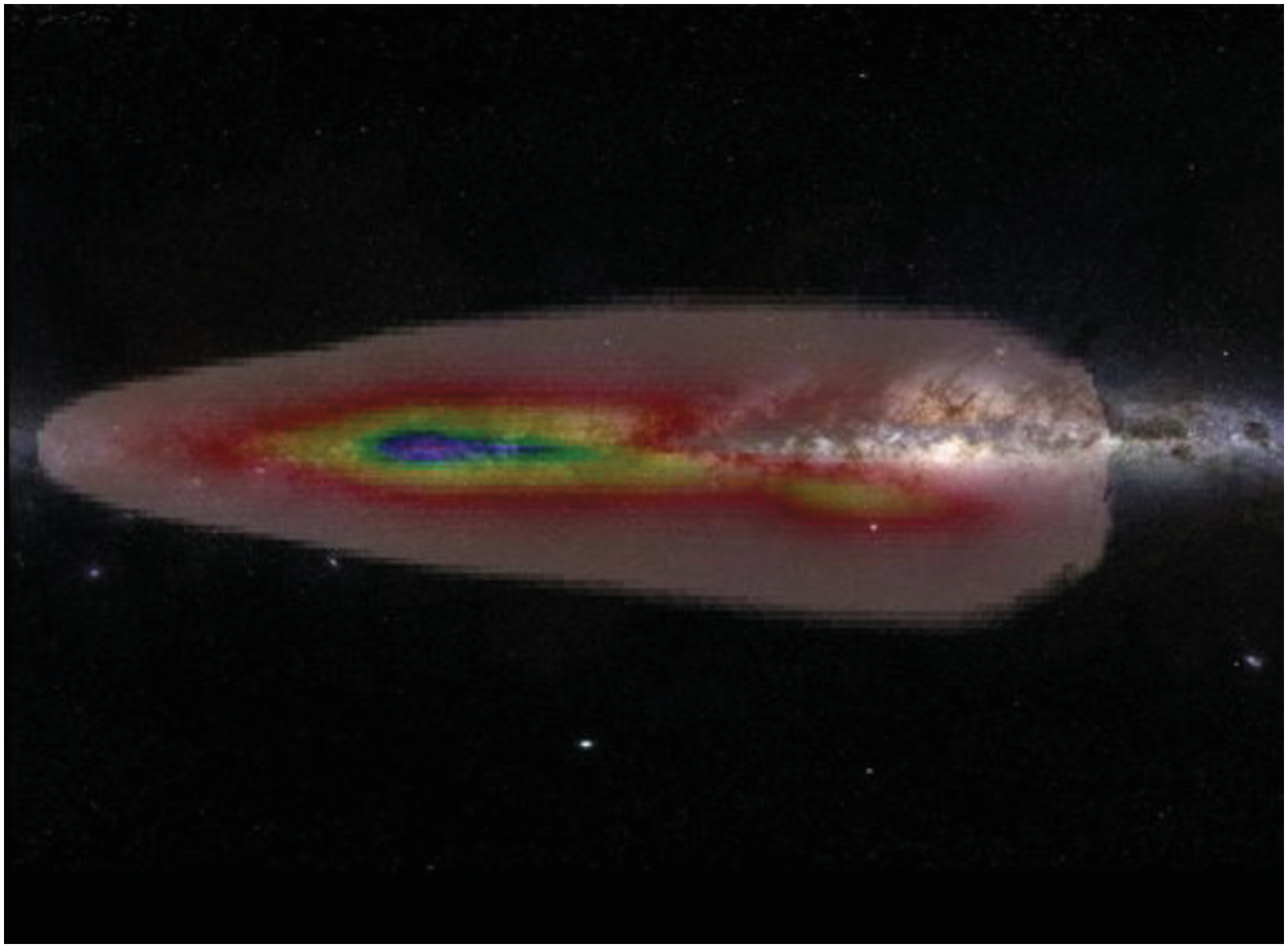}      
  \caption{These images show the expected 3D distribution of the contents of the Gaia catalogue in the Milky Way. A simulation of the Gaia catalogue is overlaid to an artistic top view of our Galaxy (left, NASA/JPL-Caltech/R. Hurt) and to an illustration of a side view of the Galaxy (right, Gigagalaxy zoom, ESO/S. Brunier/S. Guisard: the Milky Way as seen from ESO, Chile). 
             The colours of the overlaid simulations show the expected density of the one-billion 
           stars in the catalogue in different regions of the Milky Way, ranging from purple-blue 
           very high densities around the Sun to pink low densities farther from it. The `spikes' 
           pointing away from the Sun are due to windows in the interstellar extinction, allowing 
           deeper observations. Notice in particular the region in yellow and red, just below the 
           galactic center. It corresponds to part of the high-density bulge visible through an extinction 
           window around the galactic central region.} 
  \label{luri:fig2}
\end{figure*}

\subsection{Spatial distribution}

These simulations show that the Gaia catalogue will sample a large fraction
of the galactic volume, thoroughly mapping the solar neighbourhood, providing 
large numbers of objects for a substantial part of the disk and reaching
the central parts of the Galaxy although not the center itself. All components of the Galaxy will be extensively measured. In Figure~\ref{luri:fig2}, the sampling of the Galaxy is depicted based on the 
simulation results (see \url{http://www.rssd.esa.int/index.php?project=GAIA&page=picture_of_the_week&pow=141}
for the full-resolution images).

\subsection{The Hertzsprung-Russell  diagram}

\begin{figure*}[ht!]
 \centering
\includegraphics[scale=0.55]{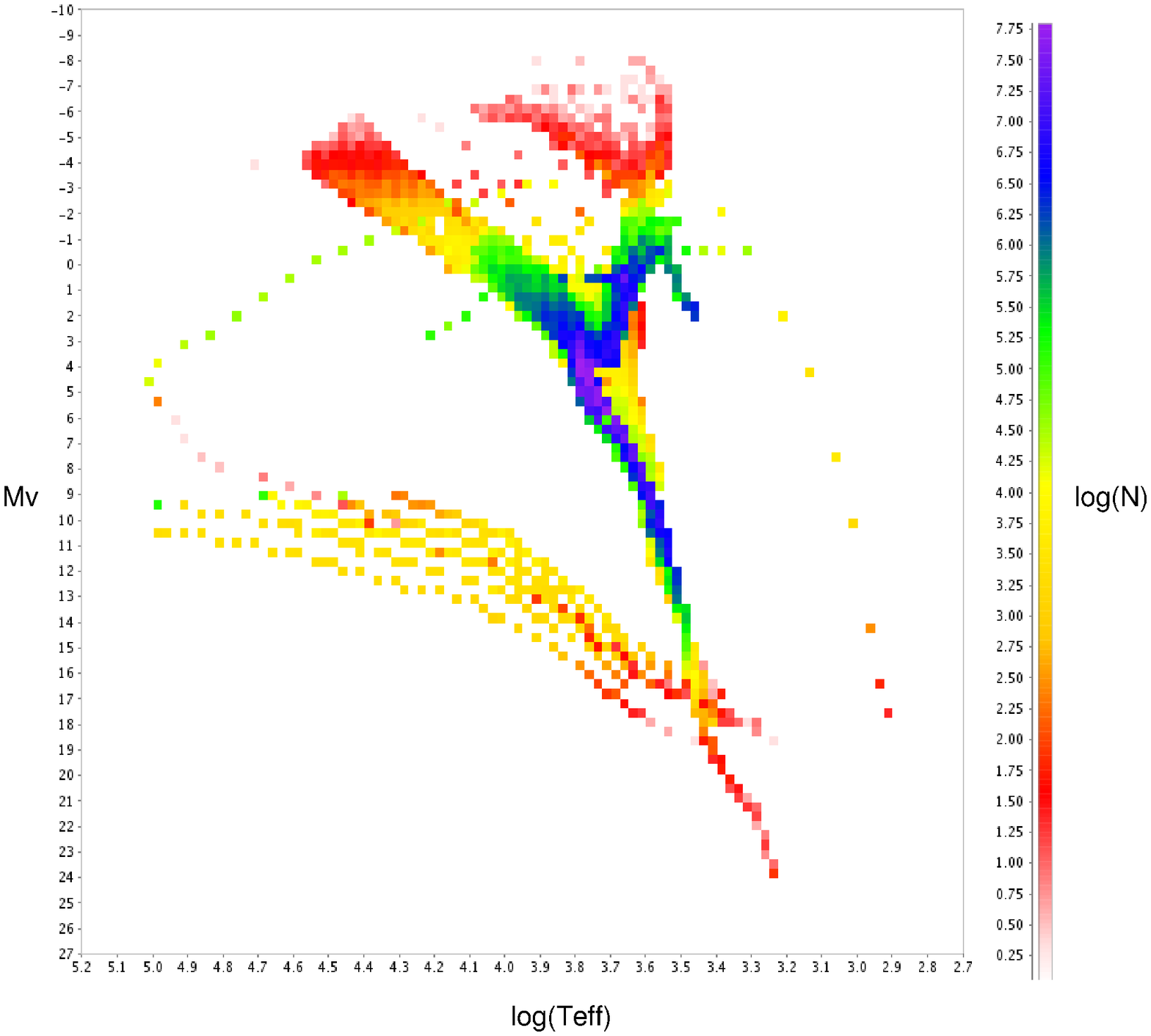}
   \caption{The Gaia HR diagram. The figure shows the expected density of catalogue objects 
           in the different regions of the HR diagram (single stars and components
           of systems alike). The colour scale gives the decimal logarithm of the number of objects
           for each [0.025K x 0.37mag] box in the diagram. To properly interpret this figure, it is important to take into account that the logarithmic scale strongly enhances the visibility of low-density areas that will be represented in the Gaia catalogue, which makes it somewhat unfamiliar compared with the usual HR diagrams.}
  \label{luri:fig3}
\end{figure*}

On the other hand, the stars in the catalogue will also fully cover the
HR diagram, sampling even the regions with the rarest types of objects. 
This coverage is depicted in Figure~\ref{luri:fig3}; the densest regions
contain tens of millions of objects and even the rarest types (bottom of
the main sequence, brightest giants) are represented with some hundreds of 
objects. 

\subsection{Catalogue simulation}

\begin{figure}[ht!]
 \centering
\includegraphics[width=\columnwidth]{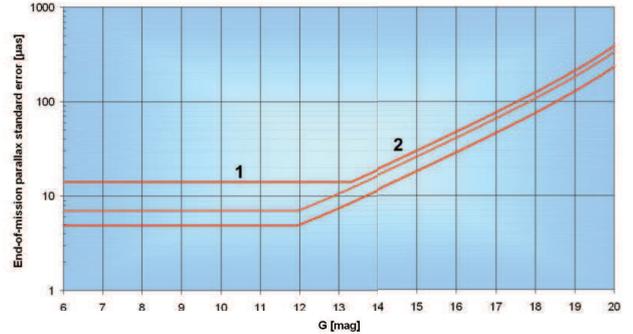}
  \caption{Sky-averaged end-of-mission parallax standard error, in $\mu$as,  as a function of the apparent Gaia G magnitude for an unreddened G2V star (V-I = 0.75 mag, V-G = 0.16 mag). The upper and lower curves reflect the range resulting from varying the sky position, the V-I colour index, and the bright-star  observing strategy. Part 1 of the curves: for $6 < G < 12$: bright star regime with CCD saturation. Part~2 of the curves: for $12 < G < 20$, photon noise regime with sky background noise and electronic noise setting in around $G~\sim$ 20 (courtesy J. de Bruijne, ESA)}
  \label{luri:fig1}
\end{figure}

The Gaia simulator also includes the capability of simulating the final mission
product: a catalogue statistically equivalent to the final data provided by 
the real mission. The expected errors on the
observable parameters (i.e. the parallax) are included in the 
simulator, statistically reproducing the expected errors in 
the final data. Figure \ref{luri:fig1} shows the sky-averaged end-of-mission parallax standard error, in $\mu$as,  as a function of the apparent Gaia G magnitude for an unreddened G2V star (V-I = 0.75 mag, V-G = 0.16 mag).

\section{Field stars}
Hipparcos provided a dramatic increase, qualitatively and quantitatively, of the basic distance information available from trigonometric parallaxes: in the '90s, there were about a thousand stars with a relative precision on parallaxes better than 10\%,  they are 22\,396 in the original Hipparcos Catalogue \citep{Perryman97a, Perryman97b} and 30\,579 in the re-reduction \citep{vanLeeuwen05,vanLeeuwen07a,vanLeeuwen07b} which used an improved and very detailed satellite attitude reconstruction, leading to an improved accuracy for stars brighter than 9$^{th}$ magnitude. 

Before the publication of the Hipparcos results, all published trigonometric parallaxes obtained from ground-based telescopes were collected in the Fourth Edition of the General Catalogue of Trigonometric Stellar Parallaxes \citep{vanAltena95}. It contains 15\,994 parallaxes for 8\,112 stars published before the end of 1995, i.e. 1722 new stars with respect to the previous edition  \citep{Jenkins63}, all reduced  to the same {\it system} of weighted absolute parallaxes. The mode of the parallax accuracy for the newly added stars (0.004\arcsec s.e.) is considerably better than in the previous editions (about 0.016\arcsec), showing the progress made over 30 years on these measurements: the use of CCD detectors greatly improves the accuracy and limiting magnitude over photographic plates.  

\begin{figure*}[ht!]
 \centering
\includegraphics[width=0.271\textwidth,clip]{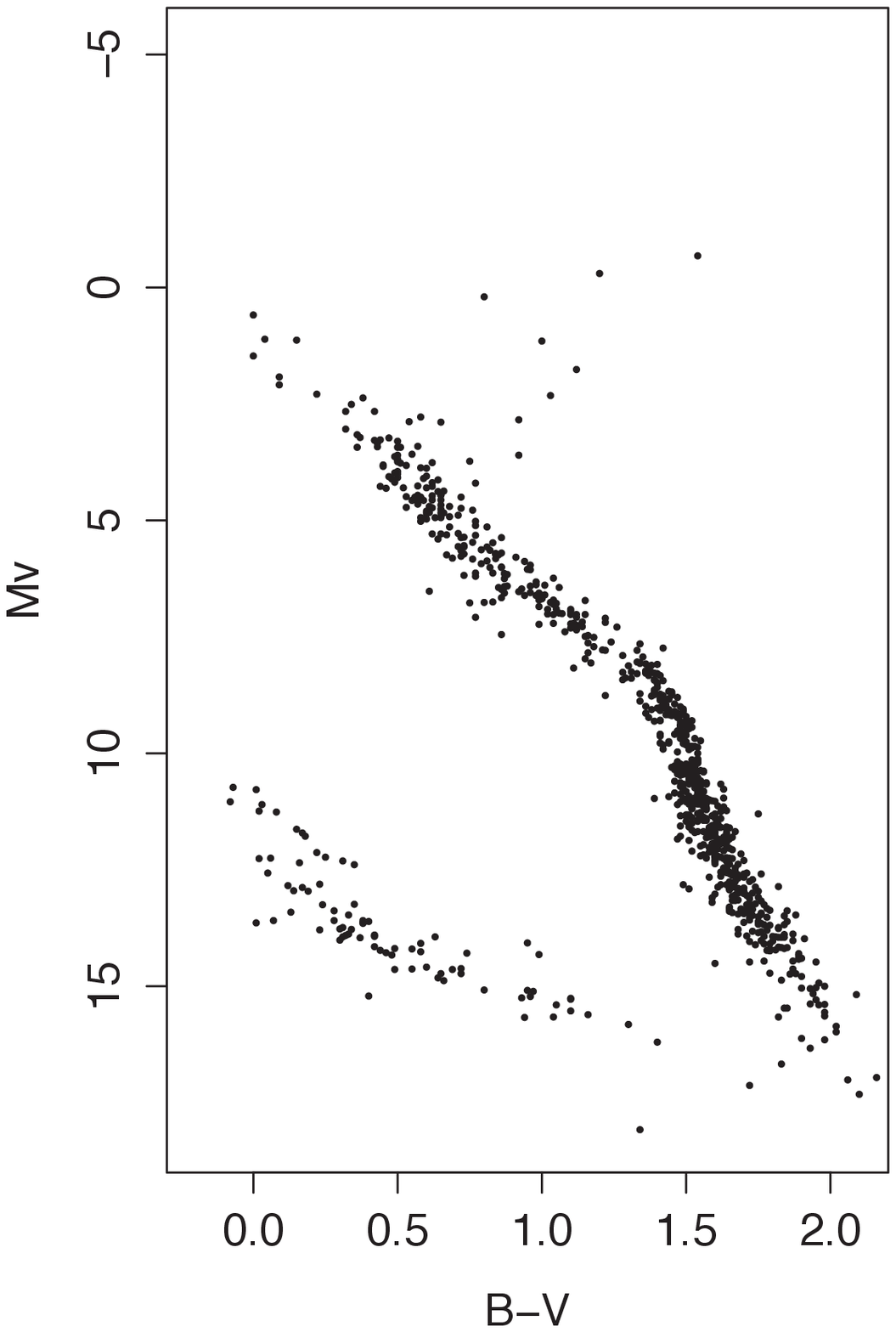}
\includegraphics[width=0.34\textwidth,clip]{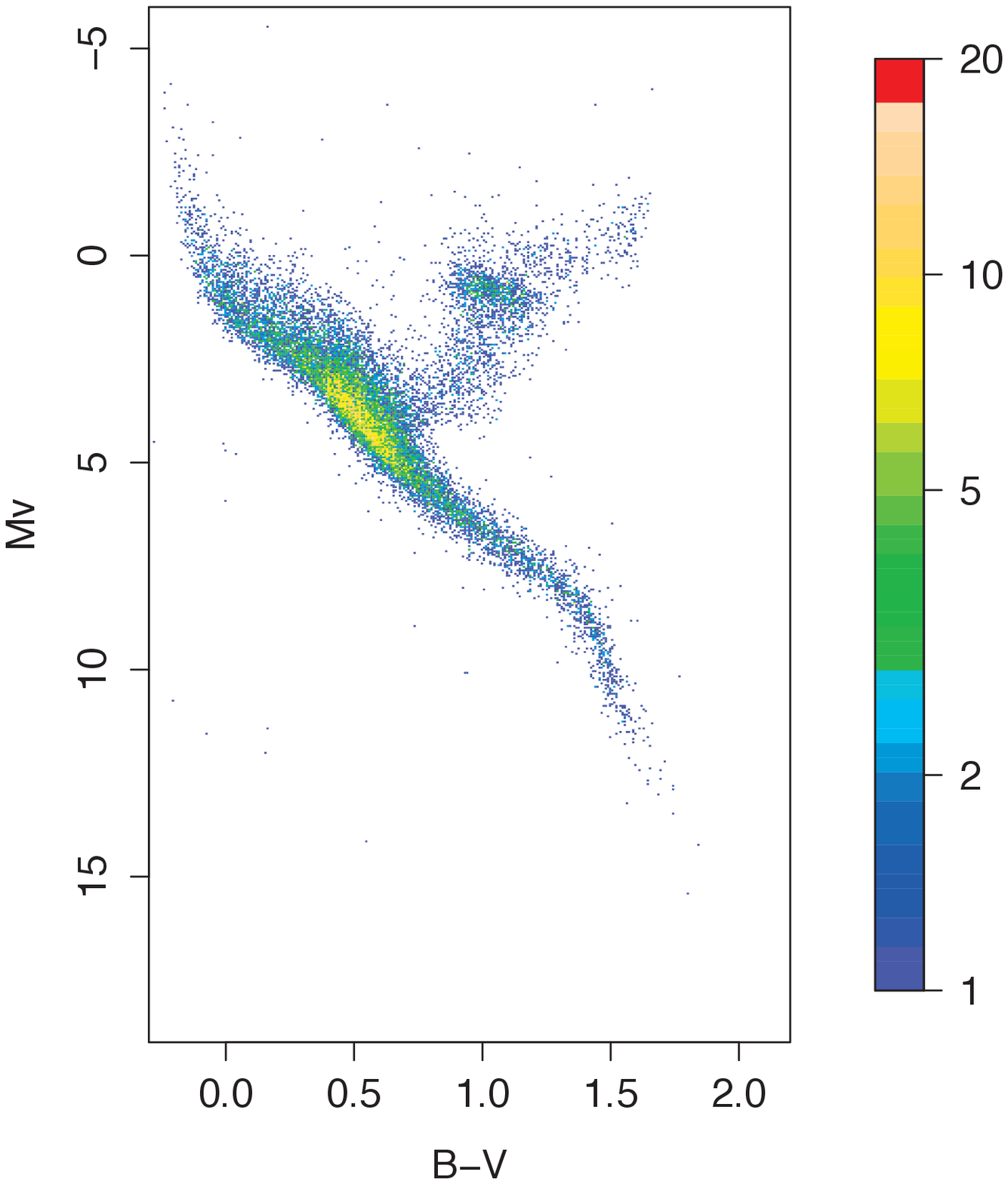}
\includegraphics[width=0.345\textwidth,clip]{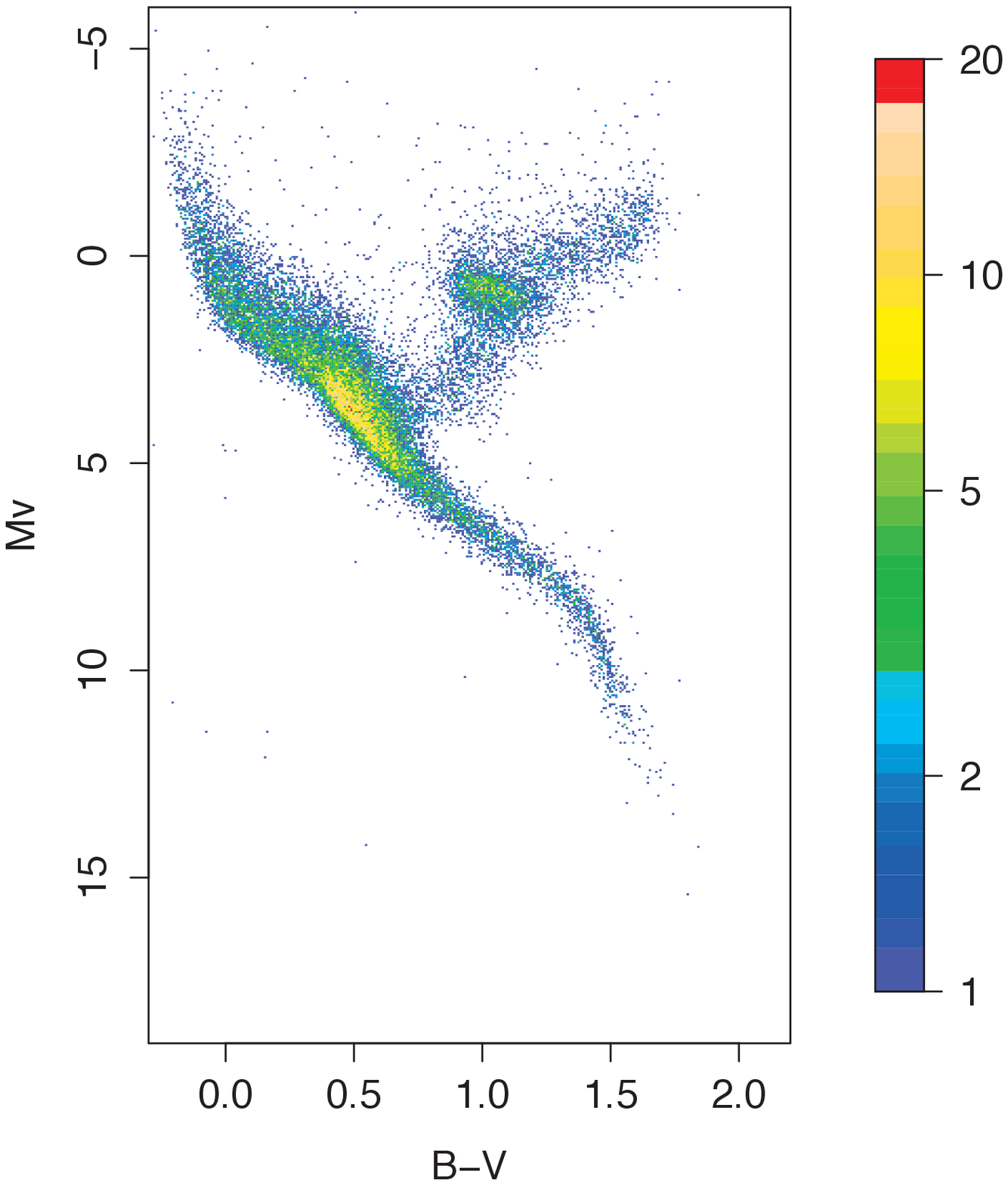}
\caption{HR diagrams drawn from the Catalogue of Nearby Stars (CNS3, left), from the original Hipparcos Catalogue (middle) and from the Hipparcos re-reduction (right). Only single stars with relative error on parallax smaller than 10\,\% and error on $B-V$ smaller than 0.025\,mag have been included. For the Hipparcos catalogue, the stellar density per cell of 0.01~mag in B-V and 0.05 in Mv is colour coded as indicated in the right scale \citep[adpated and updated from Figure~2 of][]{Turon99b}}
\label{HRHippCNS3}
\end{figure*}

The spectral type and luminosity class ranges covered by accurate Hipparcos parallaxes is considerably enlarged with respect to ground-based observations, especially towards the upper part of the main sequence and towards the giant branch, where the {\it clump} is clearly evident \citep{Perryman95, Turon99b}, and still reenforced when the Hipparcos re-reduction is considered \citep{vanLeeuwen07b}.  The main sequence is broad, up to two magnitudes, a result of undetected binaries and of a mixture of stars with various metallicities, rotation velocities, or evolutionary states. On the contrary, the faintest part of the diagram (white and red dwarf stars) is under-represented in the Hipparcos Catalogue due to the observing limitations of the satellite.  This is illustrated in Figure \ref{HRHippCNS3} where the HR diagram drawn from the Catalogue of Nearby Stars \citep[CNS3, stars estimated to be closer than 25 pc from the Sun,][left]{Gliese91} is compared with the diagrams drawn from the original Hipparcos Catalogue (middle) and the Hipparcos re-reduction (right).

\begin{figure}[h]
\includegraphics[width=\columnwidth]{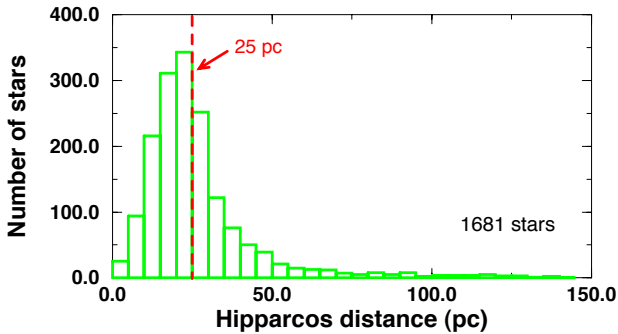}
\caption{Hipparcos distances for stars supposed to be within 25 pc of the Sun from the best ground-based trigonometric, spectroscopic and photometric parallaxes available in the 90's, compiled in the Catalogue of Nearby Stars  \citep[Figure~3 from][]{Turon99b}}
\label{distCNS3}
\end{figure}

\begin{figure}[h]
\includegraphics[width=\columnwidth]{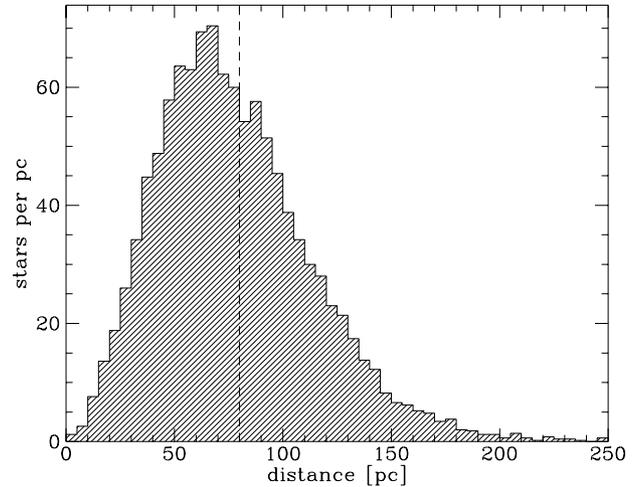}
\centering
\caption{Distance distribution of 5610 stars classified as dwarf stars in the Michigan Spectral Survey and estimated to be at distances $\leq$ 80 pc. 2384 are shown by Hipparcos to be at distances $>$~80~pc  \citep[Figure~1 from][]{Binney97}}
\label{Binneyfig1}
\end{figure}

Moreover, as shown in Figures \ref{distCNS3} and \ref{Binneyfig1}, comparisons of Hipparcos distances with ground-based distance determinations show that many distances were previously underestimated and that, even for stars in the very close neighbourhood of the Sun, distances were very poorly known: some 40\% of good quality CNS3 stars, with distances mainly obtained from trigonometric parallaxes are farther (and even in some cases much farther) than 25~pc, some 40\% of 5610 stars classified as dwarf stars in the Michigan Spectral Survey \citep{MSS}, and estimated to be at distances smaller than 80~pc from spectroscopic distance determinations, are shown by Hipparcos to be at distances larger than 80~pc \citep{Binney97}. 

As we have seen in Section~\ref{simulations}, what is expected from Gaia is still many orders of magnitude further: a systematic survey down to magnitude 20, an extreme astrometric accuracy and simultaneous astrophysical characterisation of the observed objects by multi-epoch spectrophotometric and radial-velocity measurements. The predicted end-of-mission parallax standard errors, averaged over the sky, for unreddened stars vary from 5-14 $\mu$as for the brightest stars ($6< V < 12$ mag.), through 9 to 26 $\mu$as for stars at magnitude 15 depending of their colour, to 100-330~$\mu$as at the faint end of the programme \citep[see Figure~\ref{luri:fig1}, and][]{deBruijne11a}. Hipparcos has provided a solid basis for distance measurements in the Solar neighbourhood, essentially up to 150~pc, mainly for thin disk stars, with few thick disk or halo stars and none in the bulge. Gaia will survey all stellar populations in the close-side of the Galaxy, from some less obscured windows towards the center up to the edges of the Milky Way, and can reach up to 15 kpc with relative accuracies better than 10\% on the parallax for the brightest stars, and decreasing performance with increasing magnitude and extinction in the line-of-sight. These exquisite performances will, of course, have a major impact on luminosity calibrations as shown in Table~\ref{Lum_calib}. 

\begin{figure*}[ht!]
\centering
\includegraphics[scale=0.95]{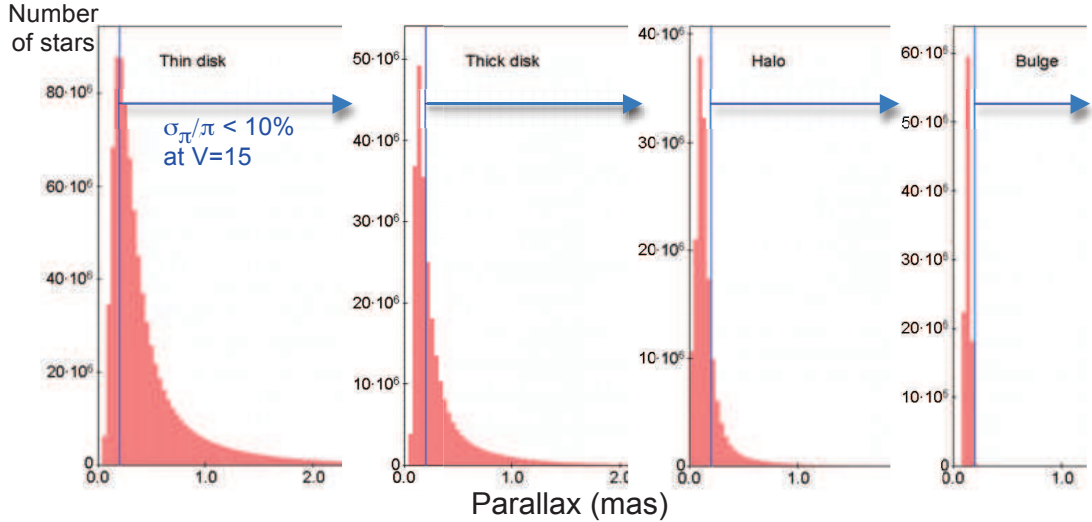}
\caption{Expected sampling of galactic stellar populations with Gaia: parallax distributions for the thin and thick disks, the halo and bulge. In these histograms, the parallaxes are given in mas in the x-axis, while the number of stars in each bin is given in the y axis. The relative error on parallaxes for all stars brighter than $V = 15$  will be smaller than 10\% for all stars situated at the right of the blue lines}
\label{Pop_sampling_simu}
\end{figure*}

\begin{figure}[ht!]
\centering
\includegraphics[width=\columnwidth]{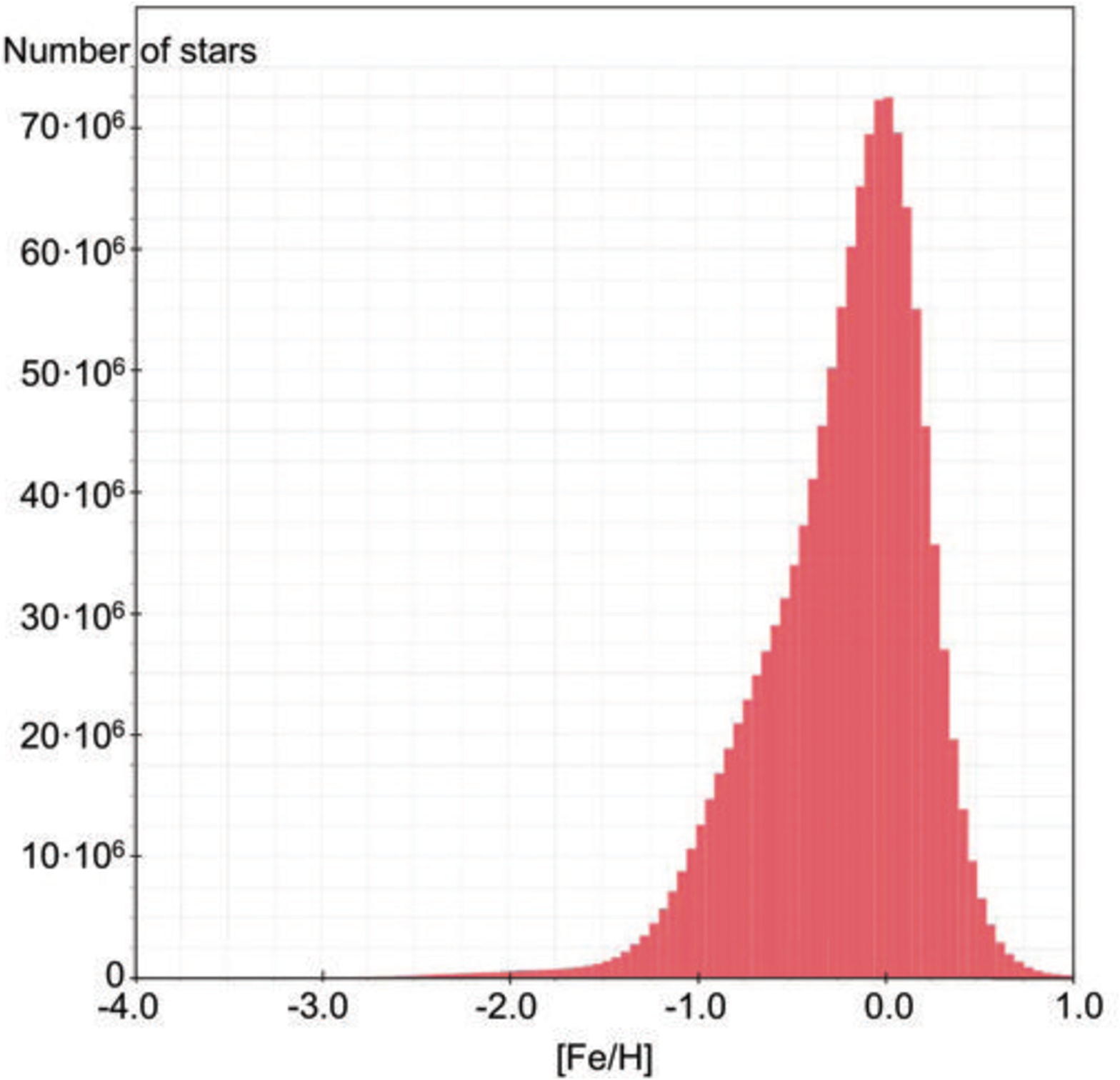}
\caption{Expected sampling in [Fe/H]}
\label{Stars_FeH_Histogram}
\end{figure}

The expected sampling of the major galactic stellar populations is illustrated in Figure~\ref{Pop_sampling_simu}: for all the stars situated at the right of the blue lines in each of the four histograms, i.e. a very significant fraction of all observed stars in each of the population, the relative error on parallaxes for all stars brighter than $V = 15$  will be smaller than 10\%. The expected sampling in [Fe/H] is presented in Figure \ref{Stars_FeH_Histogram}. This huge and systematic sampling of stars of all metallicities, including the systematic detection of metal poor and extremely metal poor stars, is absolutely crucial for main sequence fitting and for the calibration of abundance effects on luminosity calibrations of the various categories of stellar candles.

The estimations given in Table~\ref{Lum_calib} and the histograms of Figures~\ref{Pop_sampling_simu} and \ref{Stars_FeH_Histogram} are based on simulations of the Gaia mission performed within the frame of the Gaia DPAC consortium \citep{Robin11,Luri11,Czekaj11,Robin09} and make use of the simulator described in Section~\ref{simulations}.

\begin{table*}
\caption{Luminosity calibrations: from Hipparcos to Gaia \citep[adapted from][and updated] {TuronPerryman99}}
\label{Lum_calib}
\begin{center}
\small
\begin{tabular}{lccl}
\tableline 
                               &             &                    &\\ [-0.2 cm]
                               & Hipparcos (\tablenotemark{a})  &  Hipparcos   &~\,~\,~\,~\,~\,GAIA \\ 
                               &             &  re-reduction (\tablenotemark{b})  &   \\ 
                               &             &                    &\\ [-0.2 cm]
\tableline
                               &             &             &  \\ [-0.2 cm]
$\sigma_{\pi}/\pi <$ 0.1 \%  &  -  &   3    & $\sim$ 100\,000 stars \\
                               &             &             &  \\ [-0.2 cm]
$\sigma_{\pi}/\pi <$ 1 \%  &  442 stars  &  719 stars  &$\sim11\times10^{6}$ stars \\
                               &             &              &up to  1\,~\,\,--~\,\,2 kpc (Mv $<$ 0) \\
                               &             &              &up to  0.5\,--~\,1 kpc (Mv $<$ 5) \\
                               &             &              & \\ [-0.2 cm]
$\sigma_{\pi}/\pi <$ 10 \%     & 22\,396 stars & 30\,550 stars &  $\sim150\times10^{6}$ stars \\
                               &             &              &up to 10\,--\,15 kpc (Mv $<$ --5) \\
                               &             &              &up to ~\,7\,--\,10 kpc (Mv $<$ 0) \\
                               &             &              &up to ~\,2\,--~\,\,3 kpc (Mv $<$ 5) \\
                               &             &               &\\ [-0.2 cm]
                               \tableline
                               &             &               &\\ [-0.2 cm]
Error on Mv (V = 10) ~\,~\,  & \multicolumn{2}{c}{0.5 mag at 100 pc} & 0.002\,-\,0.007 mag  at 100 pc   \\
\quad due to error on $\pi$ &  &                                                          & 0.2\,-\,0.7 mag at 10 kpc \\
                               &             &              & \\ [-0.2 cm]
Stellar populations  & \multicolumn{2}{c}{mainly disk} & all populations, even the rarest \\               
                               &             &              & \\ [-0.2 cm]
HR diagram $<$ 10\% & \multicolumn{2}{c}{-5.5 $< Mv <$ 14} &  all magnitudes  \\          
                               &    \multicolumn{2}{c}{-0.3 $< B-V <$ 1.9}   &  all colours   \\
                               &             &               &\\ [-0.2 cm]
                               \tableline
\end{tabular}
\tablenotetext{a}{\cite{Perryman97a,Perryman97b}}
\tablenotetext{b}{\cite{vanLeeuwen05,vanLeeuwen07a,vanLeeuwen07b}}
\end{center}
\end{table*}

\section{Open clusters}
Nearby open clusters, and especially the Hyades, are essential first steps in the establishment of the fundamental cosmic distance scale. Hipparcos offered the first opportunity of determining accurate distances to individual stars in nearby open clusters, then making possible the determination of the mean distances of these core members without any assumption about their chemical composition or reddening, and without any use of a stellar model. Moreover, membership determination is a critical issue in which the Hipparcos distance and proper motion accuracy proved to be crucial. 

Hipparcos observed stars in all open clusters closer than 300~pc and in the richest clusters up to about 500~pc. A careful selection was made, relying on the membership information available in the 80's and taking into account the strong limitations imposed by the satellite observing constraints on the maximum stellar sky density \citep{Mermilliod89}. As a result, about 150~stars were observed in the closest open cluster, the Hyades, 50 in the Pleiades and between 10 and 40 for the other clusters closer than 300~pc. For further clusters, the number of observed stars varies from 5 to 10.

\subsection{The Hyades}
Very accurate individual distances and proper motions were obtained from Hipparcos results for about 150~stars within 15~pc of the Hyades cluster centre, including a dozen of new candidate members. These led to the first three-dimensional description of the cluster and to a very accurate determination of the distance of the observed centre of mass for the objects observed by Hipparcos within 10~pc of the cluster centre: D= 46.34 $\pm$ 0.27 pc, corresponding to a distance modulus $m\,-\,M$ = 3.33 $\pm$ 0.01 mag \citep{Perryman98}. These values have been confirmed in the Hipparcos re-reduction which led to a distance modulus of $m\,-\,M$ = 3.334 $\pm$ 0.0024 mag for 150 cluster members within 15~pc from the center \citep{vanLeeuwen09}. 

\begin{figure}[h]
\centering
\includegraphics[width=\columnwidth]{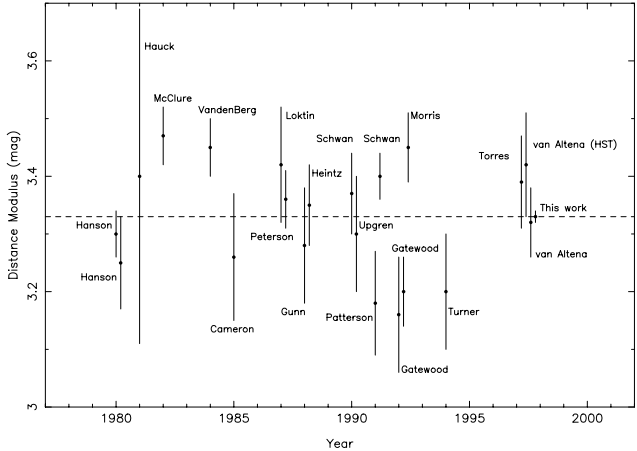}
\caption{Hyades distance modulus in 1998 \citep[Figure 1 from][]{Perryman98}}
\label{Hyades_dist_mod}
\end{figure}

\begin{figure}[h]
\centering
\includegraphics[width=\columnwidth]{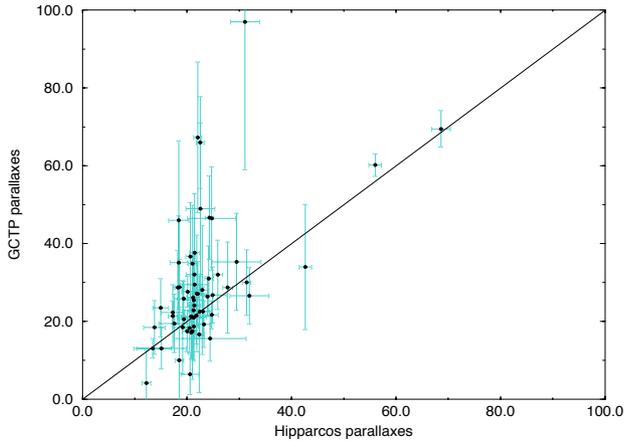}
\caption{A comparison between the GCTSP parallaxes  \citep{vanAltena95} and the Hipparcos parallaxes (both in mas) for the 60 candidate members in common between the two catalogues \citep[Figure 14 from][]{Perryman98}}
\label{Hyades_GCTSP}
\end{figure}

The remarkable improvement obtained with respect to earlier measurements is illustrated in Figure \ref{Hyades_dist_mod}. This improvement is easily understandable when comparing the individual accuracies of trigonometric parallaxes (which remain the reference for the calibration of other methods) for stars in common between Hipparcos and the General Catalogue of Trigonometric Stellar Parallaxes as given in Figure~\ref{Hyades_GCTSP}. These Hipparcos individual parallaxes are in excellent agreement with those obtained recently using the Hubble Space Telescope \citep{McArthur11}.

This strong constraint on the Hyades distance, combined with a much more reliable information on cluster membership and the very accurate Hipparcos proper motions have been used for many further kinematic and dynamical studies showing for example that the internal velocity dispersion is 0.3 kms$^{-1}$ and that the mass segregation is clearly visible \citep{Perryman98,vanLeeuwen07b,Perryman09}. Moreover, the characteristics of the cluster HR diagram have been extensively compared with stellar models computed for the Hyades metallicity, [Fe/H] = 0.14 $\pm$ 0.05, in order to determine the helium content and age of the cluster: Y = 0.26 $\pm$ 0.02 and 625 $\pm$ 50 Myr for \cite{Perryman98}, Y = 0.0.255 $\pm$ 0.009 and an upper limit of $\sim$650~Myr for \cite{Lebreton11}.

Gaia will provide a much deeper investigation of the cluster, with a systematic survey of membership down to magnitude $V=20$. This is to be compared with the Hipparcos observability limit ($V \sim 12$~mag) that led to a progressive incompleteness of the sampling with increasing magnitude, which constitutes a potential source of bias in the determination of the mean cluster distance. In addition, the Hipparcos accuracy rapidly decreases with increasing magnitudes, resulting in an increased scatter for the faint end of the HR diagram. On the contrary, Gaia will extensively cover the faint end of the cluster luminosity function down to absolute magnitudes 16-17, i.e. down to the faintest red (down to M8) and white dwarf stars, providing an overall space and mass density distribution from the center to the outer regions of the cluster and a complete picture of the extent and depth of the Hyades. 

Moreover, Gaia will also provide, in addition to astrometric parameters of extreme accuracy, a systematic detection of double or multiple systems and a complete and homogeneous set of photometric data. Finally, spectroscopic data for the brightest part of the cluster will be obtained by the Radial Velocity Spectrometer aboard Gaia down to apparent magnitudes 16-17, including all Hipparcos targets and many more. These observations will allow the determination of radial velocities, effective temperatures, rotational velocities, metallicities and abundances of a number of elements. A very clean observational HR diagram will thus been obtained and comparison with stellar models through theoretical isochrones will lead to much more reliable helium abundance and age determinations.

\subsection{Other nearby open clusters}
For open clusters closer than about 300~pc, the accuracies on the mean distances derived from Hipparcos parallaxes range from 0.2 to 0.3~mas for the Hipparcos Catalogue \citep{Robichon99, vanLeeuwen99} and from 0.1 to 0.2~mas for the Hipparcos  re-reduction \citep{vanLeeuwen09}. The error increases to 0.2 to 0.5~mas for clusters up to 500~pc. These mean distances were generally in good agreement with earlier ground-based estimates.

However, the mean distance obtained for the Pleiades \citep{Robichon99, vanLeeuwen99} was significantly smaller than the distance obtained from main-sequence fitting and led to interesting controversy \citep[see][and references herein]{Perryman09,vanLeeuwen09}. Various efforts have been made to identify causes of this discrepancy: systematics in the Hipparcos parallaxes, depth of the cluster, differences in the star samples (limiting magnitude induces that, at the distance of the Pleiades, only stars bluer than $B-V \sim 0.5$ are observed by Hipparcos while main sequence fitting rather considers redder stars), errors in He abundance or metallicity, age or reddening. So far all these attempts to explain the discrepancy have been unsuccessful.

\begin{figure}[t]
\centering
\includegraphics[width=\columnwidth]{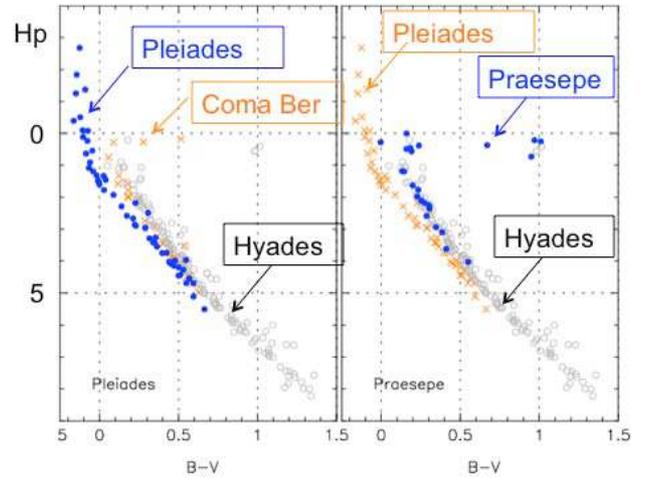}
\caption{HR diagrams for Pleiades (left), and Praesepe (right), shown as blue full dots in each diagram, and respectively compared with the Hyades and Coma Ber and with Hyades and Pleiades stars \citep[adapted from Figure 11 of][]{vanLeeuwen09}}
\label{vanL_2009_fig11}
\end{figure}

The global Hipparcos re-reduction by \cite{vanLeeuwen07a} led to a significant reduction in the error correlation levels in the Hipparcos astrometric data and to an improved accuracy on the individual distances of the bright stars observed in the Pleiades and in the other nearby clusters \citep{vanLeeuwen09}. The resulting value for the Pleiades mean distance, 120.2 $\pm$ 1.9 pc, is more accurate by a factor two but still significantly smaller than the distance obtained by main-sequence fitting. However, the positions of the main sequences are consistent within groups of clusters and this leads to suspect an age-related effect (see Figure~\ref{vanL_2009_fig11}): the Hyades and Praesepe sequences overlap, the Pleiades sequence is clearly distinct with a difference in slope with respect to the Hyades-Praesepe sequence but is coinciding with clusters like NGC 2516 or Blanco~1 (and also IC 2602 or IC 2391 for the bluest part of their sequence). Coma Ber, with an age estimated to be intermediate between the Hyades and the Pleiades, has a sequence situated between the other two groups of sequences.

A final explanation for these questions will very probably have to wait for the extensive survey that will be provided by Gaia, leading to very clean sequences in the HR diagram. Hundreds to thousands of stars will be observed by Gaia in thousands of open clusters in our Galaxy, providing reliable membership and homogeneous astrometric, photometric and spectroscopic data. A complete sampling in age and chemical composition will then be available, giving a unique tool both for distance scale building and comparison with stellar models. Moreover, Gaia will even be able to observe open clusters in the Magellanic Clouds : for example, about 3000~stars with typical errors on individual parallaxes of $\sim$~200 $\mu$as will be observed in the massive cluster R136 in the Large Magellanic Cloud (LMC), providing data on clusters in a vey different environment \citep{deBruijne11b}.

Table \ref{open_clusters} summarises the major step that will be made from Hipparcos to Gaia.

\begin{table}[h]
\begin{center}
\caption{Open clusters: from Hipparcos to Gaia \citep[adapted from][and updated]{Turon99a,redbook}}
\label{open_clusters}
\footnotesize
\begin{tabular}{ll}
\tableline 
                                                    &             \\ [-0.2 cm]
                                                    &  First cluster observed in 3-D: the Hyades \\
Hipparcos                                    &  Hyades cluster with mean distance to $<$ 1~\% \\
                                                    &  6 clusters with mean distance to $<$ 5~\% \\
                                                    & 4 clusters with mean distance to 5 - 10~\% \\
                                                    & 8 clusters with mean distance to 10 - 20~\% \\
                                                    &             \\ [-0.2 cm]
\tableline 
                                                    &             \\ [-0.2 cm]
Hipparcos                                    &  8 clusters with mean distance to $<$ 3~\% \\
 re-reduction                                & 11 clusters with mean distance to $<$ 10~\% \\
                                                    &             \\ [-0.2 cm]
\tableline 
                                                    &             \\ [-0.2 cm]
                                                    & Complete membership census down to a \\
                                                    & \hspace{3 mm} limiting magnitude depending on \\
                                                    &  \hspace{3 mm} the distance of the cluster  \\
Gaia                                             & 3-D observation to $\sim$ 500-1000 pc depending   \\
                                                    & \hspace{3 mm} on age and interstellar absorption \\
                                                    & All mean distances to better than $<$ 1~\% \\
                                                    & Observation of clusters in the LMC \\
                                                    & Many new clusters to be discovered \\
                                                    &             \\ [-0.2 cm]
\tableline
\end{tabular}
\end{center}
\end{table}

\section{Globular clusters}
All globular clusters of our Galaxy are beyond the reach of accurate Hipparcos trigonometric parallaxes. However, Hipparcos data have been extensively used to revise the distances and ages of globular clusters \citep{Cacciari99}, in particular through the very accurate trigonometric parallaxes of a set of $\sim$~500 nearby field subdwarf stars used as Population~II calibrators and carefully selected to be part of the Hipparcos Input Catalogue \citep{Turon92}. This sample of metal-poor dwarf stars is considered to be almost complete, or at least adequately representative, down to $V = 10$, by \cite{Carretta00}. 

\begin{figure}[h]
\centering
\includegraphics[scale=0.7]{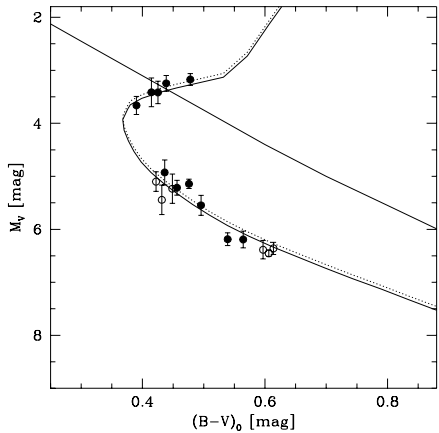}
\caption{Luminosity-colour diagram showing the M92 sequence fitted to the position of subdwarfs of similar metallicity observed by Hipparcos. The location of the evolved field subdwarfs (along the subgiant branch) provides a strong indication that M92 and the most metal-poor subdwarfs are coeval. The upper line is the Hyades main sequence. Suspected or detected binaries are shown as open circles    \citep[Figure 5a from][]{Pont98}}
\label{Pont_etal_1998_fig5a}
\end{figure}

A thorough discussion of early Hipparcos distances and ages of globular clusters obtained by subdwarf sequence fitting \citep{Reid98,Pont98,Gratton97} is given by \cite{Carretta00}, leading to systematically larger distances and smaller ages, with an average age of 12.9~$\pm$~2.9 Gyr. They also underline the various sources of uncertainty entering into this method (mainly: strong sensitivity to metal abundance, reddening, undetected binaries, correction of observational biases, stellar physics involved in the evolutionary models). An example of subdwarf sequence fitting is shown in Figure~\ref{Pont_etal_1998_fig5a}. 

This result was a crucial first element of a solution to the {\it age paradox}, i.e. that, before Hipparcos results, the ages of the oldest globular clusters were estimated to exceed the expansion age of the Universe. The other major input from Hipparcos was the revised distance of the Large Magellanic Cloud (increased by about 10\,\%) from trigonometric parallaxes of the nearest Cepheids, later confirmed with the extensive observation of Cepheids by the HST \citep{Freedman01}.

Gaia will provide a complete kinematic membership census of stars in globular clusters down to $V\,=\,20$ except in the very central zones where the stellar density will be too high, especially for the farthest clusters. About 80 of the 150 known globular clusters are at distances smaller than 10~kpc \citep{Bica06} and, for most of them, more than 1000 stars will be observed by Gaia, leading to very accurate mean distances (to better than 1\,\%) and space velocities, and very clean colour-magnitude diagrams. The magnitude of the blue horizontal branches of observed clusters will range from $V\,=\,12.5$ for the closest to $V\,=\,19$ for clusters in the Large Magellanic Cloud, and the turn-off will typically be between 18 and 20 mag (16 for the brightest). In addition, Gaia will observe many more subdwarf stars than Hipparcos, providing a much better sampling in metallicity for main sequence fitting. Table \ref{glob_clusters} summarises the major step that will be made from Hipparcos to Gaia.

\begin{table}[h]
\begin{center}
\caption{Globular clusters: from Hipparcos to Gaia \citep[adapted from][and updated]{Turon99a,redbook}}
\label{glob_clusters}
\footnotesize
\begin{tabular}{ll}
\tableline 
                        &             \\ [-0.2 cm]
                        &  No star observed in globular clusters   \\
Hipparcos        &  Fitting to nearby subdwarf sequences \\
                        &  Systematically larger distances \& smaller ages  \\
                        &             \\ [-0.2 cm]
\tableline 
                        &             \\ [-0.2 cm]
                        & Complete membership census  \\
Gaia                &  \hspace{1 cm}      except in very central zones \\
                       & Between 100 and 100\,000 stars per cluster     \\
                       & More than 5000 stars for half of the clusters \\
                       & Mean distances to $<$ 1~\% for about 80 clusters \\
                       & Mean distances to $<$ 5~\% for all clusters \\
                       &             \\ [-0.2 cm]
\tableline
\end{tabular}
\end{center}
\end{table}

\section{Pulsating variables}
Among pulsating variables, Classical Cepheids, because of their intrinsic high brightness, are the most powerful distance indicators \citep{Clementini11} and we will mainly concentrate on them in this section. Presently about 800 galactic Classical Cepheids are known, most of which within 4~kpc from the Sun \citep{Berdnikov00, Clementini11}. There are 280 Cepheids in the Hipparcos Catalogue, 248 of which are classical Cepheids. Even though most of these Cepheids are at distances larger than about 500~pc and, as a consequence, their individual parallaxes were not measured very accurately by Hipparcos, Hipparcos data have been successfully used to determine the zero-point of the period-luminosity (P-L) relation (\cite{vanLeeuwenetal07} also determined the slope). 

The first paper using Hipparcos parallaxes for these stars  \citep{Feast97} considered 223 Hipparcos Cepheids with a mean standard error on the parallaxes of  $\sim$ 1.5 mas, but only 26 of them contribute significantly to the weight of the derived zero-point of the P-L relation, $<M_V>$\,=\,2.81~log\,$P$~-~1.43, with a standard error of the zero-point of 0.10 mag. Taking into account appropriate metallicity corrections, this led to a distance modulus of 18.70\,$\pm$\,0.10 for the Large Magellanic Cloud, to be compared with the 18.50 mag of earlier determinations. Many effects were considered to explain this difference in addition to the new trigonometric parallaxes: metallicity effects, line-of-sight reddening, unrecognised binarity, undetected overtone pulsators, etc. 

\cite{Luri98} applied their Maximum-Likelihood method \citep[LM method,][]{Luri96} to Hipparcos data to derive luminosity calibrations for Classical Cepheids and RR Lyrae. Their method takes into account all available data (luminosity, proper motions, spatial distribution and associated observational errors) and the observational truncation due to the Hipparcos limitations in magnitude. Indeed, they made the analysis that \cite{Feast97} method was highly sensitive to this kind of effects. They obtained an LMC distance modulus of 18.3~$\pm$ 0.2~mag, both from Classical Cepheids and from RR Lyrae.

The Hipparcos re-reduction provided an improvement by up to a factor of 2 for many of the 240 Classical Cepheids \citep{vanLeeuwenetal07}. They used the revised parallaxes of a subset of 14 stars, combined with 10 HST parallaxes \citep{Benedict07}, to determine the slope of the P-L-C relation and the whole sample (of which about 100 contribute significantly to the derived solution) for the determination of the zero-point. They found that these relations have similar slopes in our Galaxy and in the LMC and that their zero-points lead to a metallicity corrected LMC distance modulus of 18.39 $\pm$ 0.05 mag, in exact agreement with that of \cite{Freedman10}: 18.39 $\pm$ 0.06 mag.

\begin{figure}[h]
\centering
\includegraphics[scale=0.5]{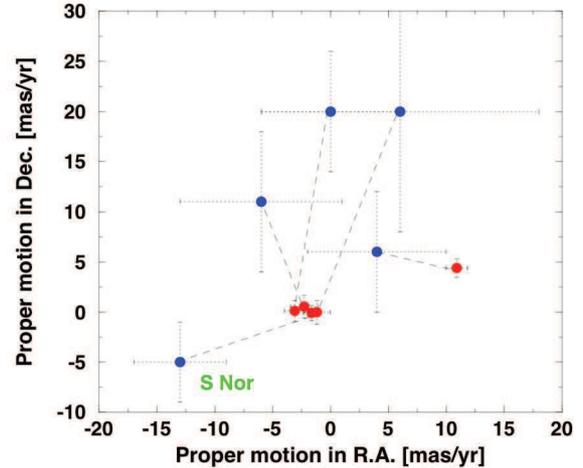}
\caption{NGC 6087 cluster membership of S Nor from Hipparcos proper motions (mas/yr). Blue dots: pre-Hipparcos proper motions. Red dots: Hipparcos proper motions  \citep[Figure 1 from][]{Lynga98}}
\label{S_Nor_Lynga}
\end{figure}

Before the availability of Hipparcos distances for Cepheids, the zero-points of the period-luminosity(-colour) relations mainly relied on Cepheids in open clusters with photometric and/or spectroscopic distances. Hipparcos proper motions also provided a significant improvement in these methods by an unambiguous determination of the membership to clusters. This is illustrated in Figure~\ref{S_Nor_Lynga} where ground-based and Hipparcos proper motions are plotted, definitively demonstrating that S~Nor is a member of NGC~6087  \citep{Lynga98}.

A crucial and probably the hottest question about Cepheids is the universality of the period-luminosity(-colour) relations. The recent literature is an illustration of this dispute with definitive as well as conflicting assertions on the question \citep{Freedman10, Sandage09, Romaniello08, vanLeeuwenetal07}. Gaia is expected to provide an extensive sampling of galactic Classical Cepheids: it will of course repeatedly observe the 800 known Cepheids, but will also discover many more. The exact number of Classical Cepheids that Gaia will observe is uncertain but will certainly be thousands: from 2000 to 8000 \citep{Eyer00, Clementini11} to $\sim$ 9000 \citep{Windmark11, Robin11}. For these galactic Cepheids, Gaia will not only bring very accurate distances, but also colours and metallicity and information about he extinction in their line-of-sight, resulting in a much improved determination of the slopes and zero-points of the galactic period-luminosity and period-luminosity-colour relations and of their dependence on metallicity. 

\begin{figure*}
\centering
\mbox{\subfigure[Hipparcos light curve of $\delta$ Cephei]{\label{delta_Ceph_Hip}\includegraphics[width=0.35\textwidth]{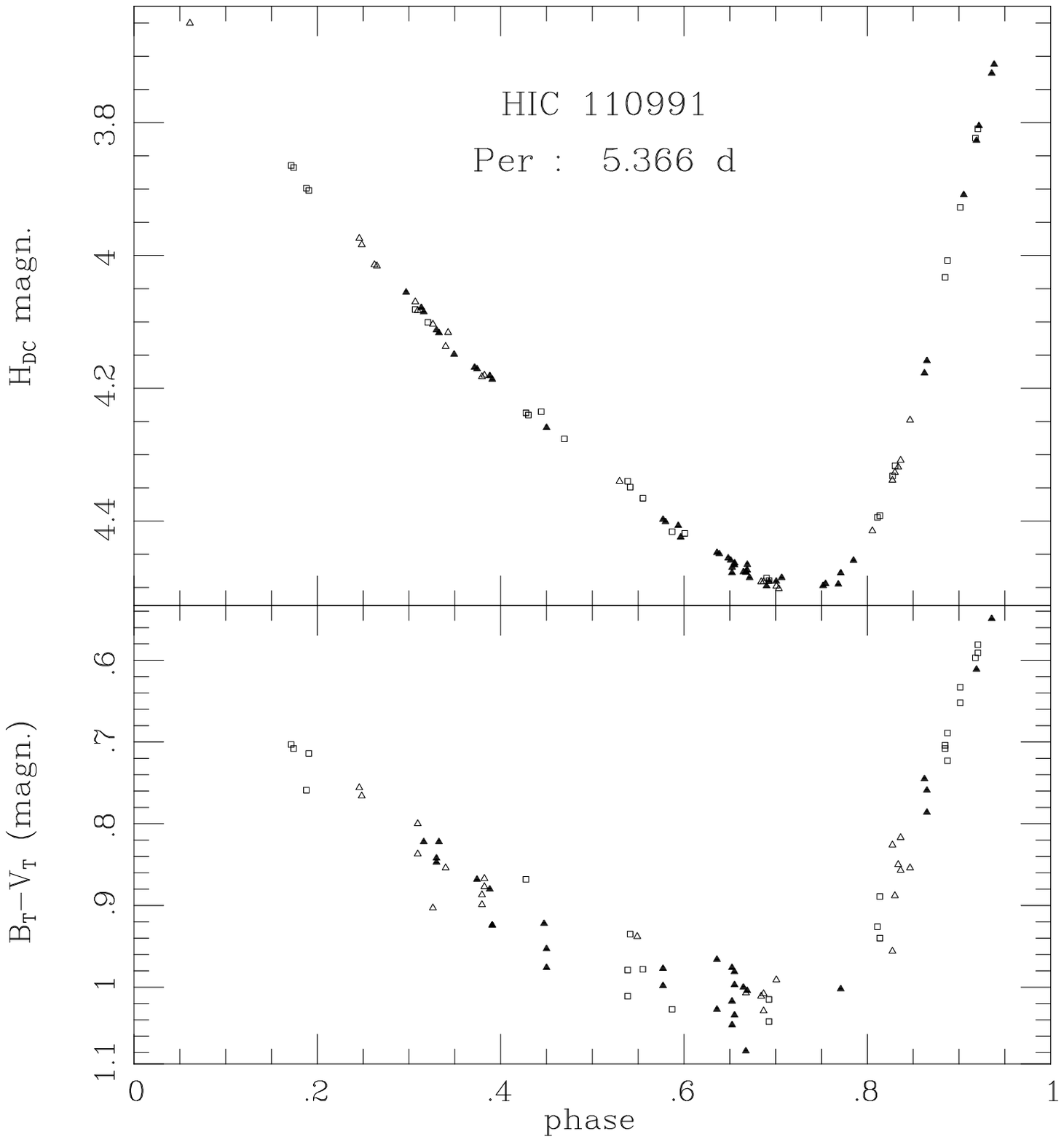}}             
\quad
\subfigure[Simulated Gaia light curve of a Cepheid in M31]{\label{Ceph_M31_Gaiar}\includegraphics[width=0.40\textwidth]{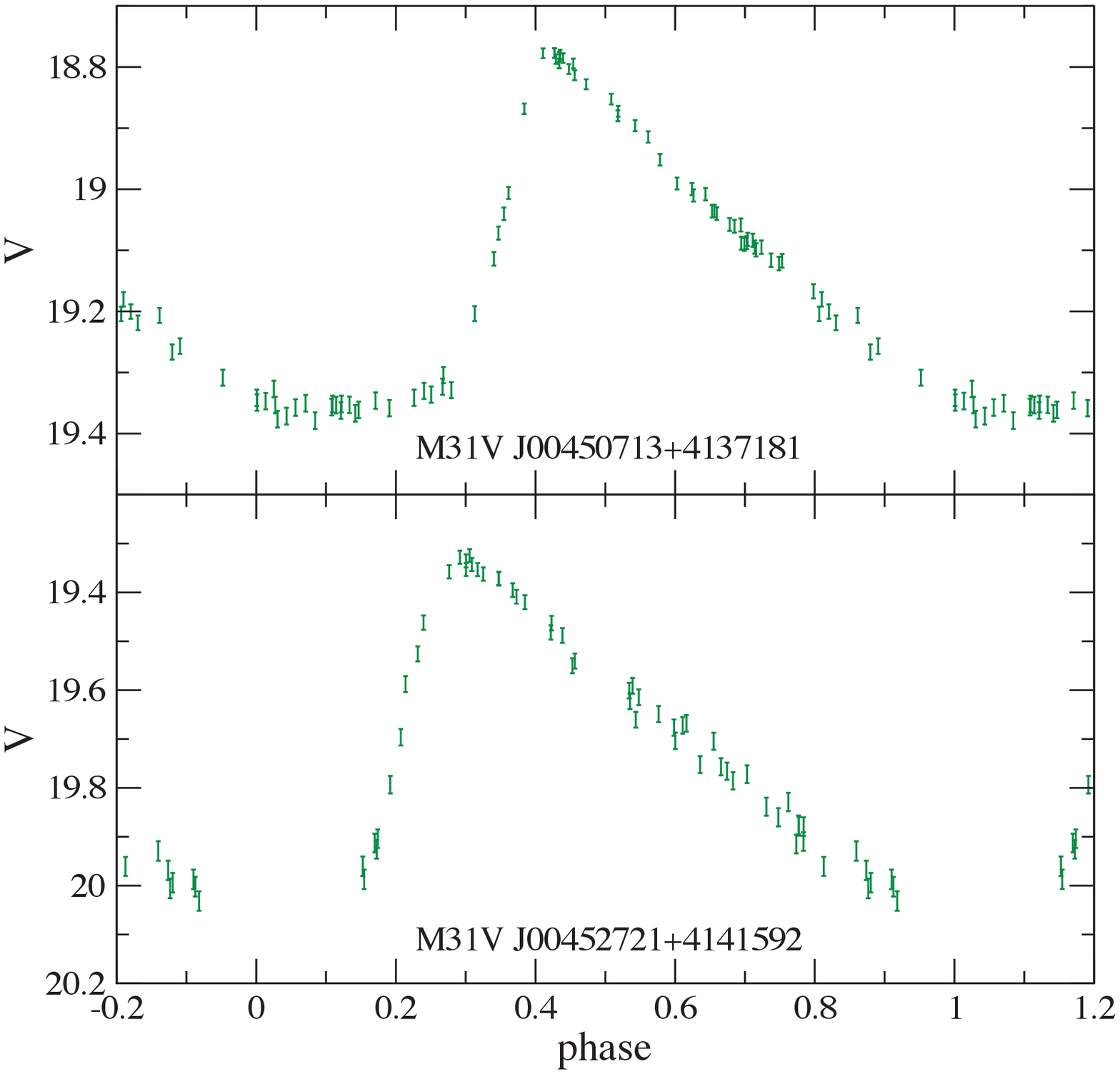}}}
\caption{Hipparcos photometry for $\delta$ Cephei compared with simulated Gaia photometry for a Cepheid in M31}
\label{Phot_Cepheids}
\end{figure*}

Figure \ref{Phot_Cepheids} is an illustration of the power of Gaia photometry and of the huge step from Hipparcos to Gaia: the Hipparcos photometry of the brightest galactic Cepheid, $\delta$ Cephei (left) is compared with the light curve expected from Gaia for Cepheids in M31 \citep[right,][]{Vilardell09}. 

In addition, 1000 to 2000 Cepheids will be observed in the Magellanic Clouds. For example, in the Large Magellanic Cloud, their $V$ magnitude will range from 12.7 to 20, leading to $\mu$arcsec accuracy for the brightest Cepheids (longest periods) down to 0.3 mas for the faintest Cepheids (shortest periods). For the bulk of them, with magnitudes between 15 and 17, Gaia astrometric accuracies will be between 20 and 80 $\mu$arcsec \citep{Soszynski08}. Of course, due to the large distances to the Clouds, no individual reliable parallaxes will be obtained, but rather mean values of the parallax for groups of stars which will allow to test the differences (or not) between the slopes and zero-points of the period-luminosity(-colour) relations in these very different environments with very different metallicities.

Table \ref{pulsating_variables} summarises the major step that will be made from Hipparcos to Gaia for pulsating variable stars. 

\begin{table*}[t]
\begin{center}
\caption{Pulsating variables: from Hipparcos to Gaia \citep[adapted from][and updated]{Turon99a,redbook}}
\label{pulsating_variables}
\footnotesize
\begin{tabular}{lll}
\tableline 
                        &          &   \\ [-0.2 cm]
                        &  \multicolumn{1}{c}{Hipparcos}  &  \multicolumn{1}{c}{Gaia}       \\ 
                        &          &   \\ [-0.2 cm]
\tableline 
                        &          &   \\ [-0.2 cm]
Classical Galactic      &   273 observed (2 new) 
                                      &  Census of galactic Cepheids with $G \leqslant 20$: $\sim$ 9000 Cepheids   \\
\hspace{0.3 cm} Cepheids     & P: 2 to 36 days   &  All periods, colours and metallicities\\
                        &   $\sim$ 30 with $\sigma_\pi / \pi \lesssim 30\,\%$ & Up to 1-2 kpc with $\sigma_\pi / \pi < 1\,\%$  \\
                        &            & All galactic Cepheids with $\sigma_\pi / \pi < 10\,\%$  \\
                        &            & Cluster membership \\
                        &            &     \\ [-0.2 cm]
\tableline 
                        &             &    \\ [-0.2 cm]
Population II    & $\sim$ 30  & $\sim$ 2000  \\
\hspace{0.3 cm} Cepheids    &             &    \\
                        &            &     \\ [-0.2 cm]
\tableline 
                        &             &    \\ [-0.2 cm]
LMC  Cepheids               &   None  & 1000-2000 Cepheids with $\sigma_\pi / \pi \sim 50-100\,\%$ \\
                        &             &   Mean distance of groups of Cepheids expected to 10\,\%   \\
                        &             &   Mean distance of LMC expected to 0.5\,\%   \\
                        &             &   Depth of LMC expected to 1\,\%   \\
                        &            &     \\ [-0.2 cm]
\tableline 
                        &             &    \\ [-0.2 cm]
RR Lyrae         &  186 observed (9 new)  & All galactic RR Lyrae: $\sim$ 70\,000 \\
                        &  Only RR Lyr with accurate $\pi$ & All metallicities  \\
                       &             &  In globular clusters: mean $\sigma_\pi / \pi < 1\,\%$        \\
                       &            &     \\ [-0.2 cm]
\tableline 
                        &             &    \\ [-0.2 cm]
All pulsating     &             & Extensive surveys of all types of variables  \\
\hspace{0.3 cm} variables &   & Astrometry, photometry and spectroscopy  \\
                       &               &   Extensive sampling versus period, colour, metallicity     \\
                       &               & Determination of the zero-points and slopes the P-L(-C) relations \\
                       &               & Determination of the intrinsic dispersion of the  P-L(-C) relations \\
                       &               & Cluster membership \\
                       &               &          \\ [-0.2 cm]
\tableline
\end{tabular}
\end{center}
\end{table*}

\section{Magellanic Clouds and Local Group Galaxies}\label{LMC-SMC}

A few dozen of stars were carefully pre-selected in the Large and Small Magellanic Clouds for Hipparcos observation, taking into account the limitations on magnitude and crowding imposed by the satellite observing mode \citep{Turon92}: 36 stars in the Large Magellanic Cloud and 11 in the Small Magellanic Cloud (SMC). At distances of about 48 and 54~kpc respectively, no direct distance determination was possible and only the proper motions were significant. However, a number of indirect methods have been developed, based (among other data) on Hipparcos observations of various types of objects: Cepheids,  Horizontal branch (HB) stars, eclipsing binaries, SN 1987a, red clump giants, Mira variables, subdwarf fitting to globular clusters, RR Lyrae \citep[see][for a review of these applications of Hipparcos data]{Perryman09}. 

In the '90s, there was still a difference between the distances obtained for the LMC from Population I indicators (Cepheids, red clump stars, Mira variables) and from Population II indicators (RR Lyrae, HB stars, subdwarf main sequence fitting), and the LMC distance was still the major source of uncertainty in the determination for the Hubble constant, $H_0$ \citep{Freedman01}. One of the first paper to reconcile these two approaches was \cite{Luri98}, who obtained an LMC distance modulus of 18.3~$\pm$0.2 mag. \cite{Feast08}, using the Hipparcos re-reduction for type II Cepheids and RR Lyrae variables obtained 18.37~$\pm$0.09 mag.

\begin{figure}[ht!]
 \centering
\includegraphics[width=\columnwidth]{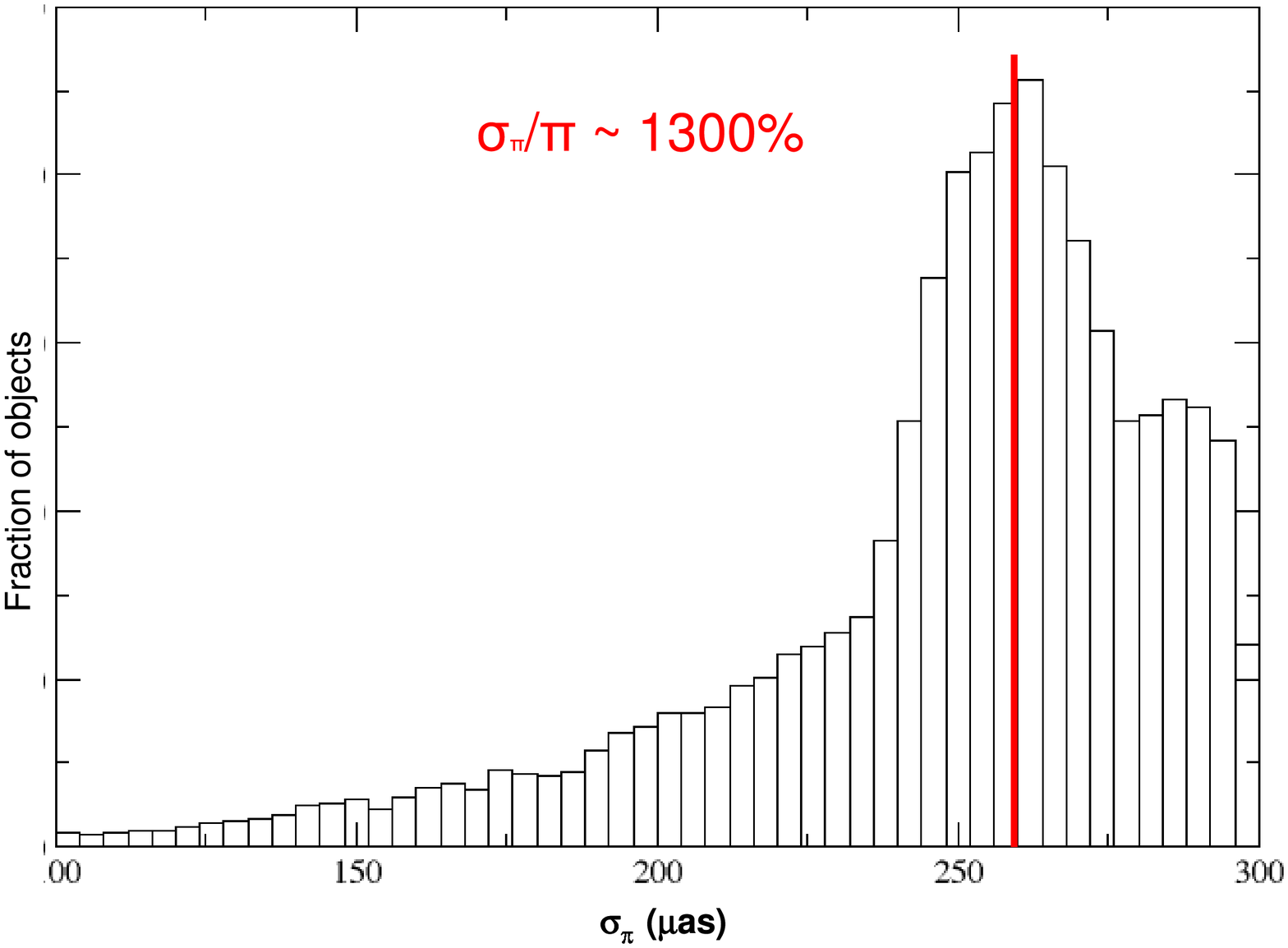}    
  \caption{Distribution of the errors in parallax for the simulated LMC objects. Notice
           that the maximum is at a relative error of 1300\%, but that there is a significant
           tail of objects reaching low relative errors.}
  \label{luri:fig4}
\end{figure}

Direct trigonometric parallaxes in the Magellanic Clouds will have to wait for Gaia which will observe millions of stars in each Cloud: $\sim$\,7\,500\,000 in the LMC and $\sim$\,1\,500\,000 in the SMC \citep{Robin11}. As mentioned above, Cepheids, but also all stars in the asymptotic giant branch, will be bright enough to be observed with Gaia astrometric accuracies from about 10 to 80~$\mu$as. However  most of the stars will have Gaia-magnitudes ($G$) of about 19-20, and thus parallax accuracies ranging from 100~$\mu$as for the reddest stars to about 300~$\mu$as for blue stars. Figure \ref{luri:fig4} shows the expected distribution of the errors in the parallaxes of the LMC objects based on the results of the Gaia simulator.

As a result, and considering that the errors on individual parallaxes are uncorrelated, the mean distances of the LMC and SMC are expected to be determined to $\sigma_\pi / \pi \approx 0.5\,\%$ for the LMC and $\sigma_\pi / \pi \approx\,1.5\,\%$ for the SMC. Moreover, the 3D distribution of various types of (giant) stars will be within reach: with errors in the mean parallaxes of $\sim$\,0.12 $\mu$as for the LMC and $\sim$\,0.24~$\mu$as for the SMC, the 3D structure of the Clouds will be observable. Indeed, with a depth of about 3000\,pc (there is still a very large uncertainty on the depths of LMC and SMC), the parallaxes vary from about 22.2 to 19.6 $\mu$as for the LMC, and from about 17.2 to 15.6 $\mu$as for the SMC. This would for example allow to check if Cepheids are particularly concentrated in the bar of the LMC. 
 
\begin{figure}[h]
\centering
\includegraphics[width=\columnwidth]{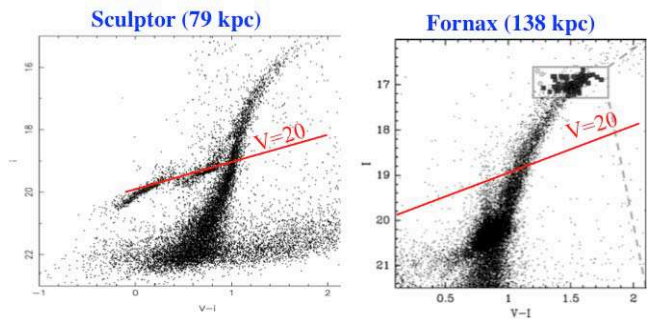}
\caption{Colour-magnitude diagrams in the Local Group (Courtesy V. Hill)}
\label{Sculptor_Fornax}
\end{figure}

As illustrated in Figure~\ref{Sculptor_Fornax} Gaia will even observe a number of individual stars in Local Group galaxies, providing an unambiguous discrimination with solar neighbourhood stars.

\section{Conclusion}
The precise knowledge of distances is an essential clue to our understanding of the structure, formation and evolution of our Galaxy and to the calibration of the luminosities of stellar candles used to estimate the distances in the Universe, very far beyond the Local Group of galaxies. Hipparcos was a major step forward, making astrometry a fundamental tool for astrophysics, as demonstrated by the quantity and quality of applications relying on its data  \citep[see][for a review of these many applications over the 10 years since the publication of the Hipparcos and Tycho Catalogues in 1997]{Perryman09}. A few of them, mainly those dealing with the Fundamental Cosmic Distance Scale, have been quoted in the present paper, with no ambition to be exhaustive but rather as illustrations of the importance of accurate astrometric data for the progress of our knowledge in this field. 

The next, spectacular, step will be Gaia, with orders of magnitude improvement in the number of observed targets (from 118\,000 for Hipparcos to 1 billion for Gaia) and in the astrometric accuracy (from 0.1~mas at magnitudes brighter than 5 to 3.0~mas at magnitude 12 for Hipparcos down to 8~$\mu$as at magnitude 12 to 300~$\mu$as at magnitude 20 for Gaia). The other, essential, characteristic of Gaia is its capability to obtain photometric and spectroscopic observations quasi-simultaneously with astrometric data. This is a key possibility for a complete study of stellar candles, especially because of the importance of metallicity effects on their luminosity. We have restricted the discussion in this paper to a few types of stellar candles, but many other potential candles will be observed among the billion of Gaia targets: red clump giants, tip the red giant branch, Mirae \citep[very promising, see][]{Whitelock11}, type II Cepheids, etc. etc. We are entering a very exciting new era for the determination of the Fundamental Cosmic Distance Scale.

\section{Acknowledgements}
{\it The authors want to acknowledge the organisers for offering them the opportunity of this review paper. 
X.Luri and E. Masana acknowledge the MICINN (Spanish Ministry of Science and Innovation) -
FEDER through grant AYA 2009-14648-C02-01 and CONSOLIDER CSD 2007 00050. The simulations
presented in this paper have been done in the supercomputer MareNostrum at Barcelona
Supercomputing Center -- Centro Nacional de Supercomputaci\'on (The Spanish National
Supercomputing Center). The authors also want to acknowledge the use of NASA's Astrophysics Data System Bibliographic Services and of the VizieR catalogue access tool, CDS, Strasbourg, France \citep{Ochsenbein00}.

\end{document}